\documentstyle[12pt]{article}

\newcommand {\e} {\mbox{\rm e}}

\newcommand {\nn}    {\nonumber}
\newcommand {\vs}[1]  { \vspace*{#1 cm} }

\newcounter{eq}
\newcounter{sc}

\newcommand {\IJMP}  {Int. J. Mod. Phys.}
\newcommand {\MPL}  {Mod. Phys. Lett.}
\newcommand {\NP}   {Nucl. Phys.}
\newcommand {\PL}   {Phys. Lett.}

\newcommand {\AP}   {Ann. of Phys.}
\newcommand {\PTP}  {Prog. Theor. Phys.}

\newcommand {\JMP}  {J. Math. Phys.}

\def\overleftrightarrow#1{\vbox{\ialign{##\crcr
 $\leftrightarrow$\crcr\noalign{\kern-1pt\nointerlineskip}
 $\hfil\displaystyle{#1}\hfil$\crcr}}}

\newlength{\minitwocolumn}
\begin{document}

\begin{flushright}
EDO-EP-42\\
June, 2001\\
\end{flushright}
\vspace{30pt}

\pagestyle{empty}
\baselineskip15pt

\begin{center}
{\large\bf Higgs Mechanism in the Presence of a Topological Term

 \vskip 1mm
}

\vspace{20mm}

Ichiro Oda

\vspace{10mm}
          Edogawa University,
          474 Komaki, Nagareyama City, Chiba 270-0198, JAPAN \\

\end{center}


\vspace{15mm}
\begin{abstract}
In cases of both abelian and nonabelian gauge groups, we study the 
Higgs mechanism in the topologically massive gauge theories in
an arbitrary space-time dimension. We show that when the conventional
Higgs potential coexists with a topological term, gauge fields become 
massive by 'eating' simultaneously both the Nambu-Goldstone boson and a 
higher-rank tensor field, and instead a new massless scalar field is 
'vomitted' in the physical spectrum.
Because of the appearance of this new massless field, the number of 
the physical degrees of freedom remains unchanged before and after 
the spontaneous symmetry breakdown.
Moreover, the fact that the new field is a physical and positive norm 
state is rigorously proved by performing the manifestly covariant 
quantization of the model in three and four dimensions. 
In the mechanism at hand, the presence of a topological term makes it
possible to shift mass of gauge fields in a nontrivial manner compared 
to the conventional value.

\vspace{15mm}

\end{abstract}

\newpage
\pagestyle{plain}
\pagenumbering{arabic}


\rm
\section{Introduction}

The Standard Model based on $SU(3)_C \times SU(2)_L \times U(1)_Y$
is a remarkable achievement. We have strong confidence that it provides
a fundamental theory of the non-gravitational interactions of quarks
and leptons valid up to energies of order 1 TeV. Moreover, it is
consistent  with the great wealth of experimental data existing at
present.

Despite its remarkable success, however, the Standard Model still 
possesses some important questions such as the multiplicity of generations,
a lot of undetermined parameters and the gauge hierarchy problem e.t.c. 
In addition to these theoretical questions, we should notice that the 
Higgs sector in the Standard Model has not been observed experimentally 
and the least understood.   
 
On the other hand, in recent years we have watched a remarkable development 
in superstring theory that
appears to be the first self-consistent theory constructed so that
all of the interactions of Nature are unified. A natural question 
then arises what modification superstring theory provides for the
Higgs sector in the Standard Model. However, superstring phenomenology,
the study of how superstring theory makes contact with physics at accessible 
energy, is still in its infancy, so we have no quantitative predictions, as 
yet, from superstring theory. Nevertheless, there are a number of important
qualitative implications and insights which have been obtained from
superstring theory. In particular, superstring theory predicts the existence 
of many of new particles such as a dilaton, an axion and perhaps other 
scalar moduli. Together with them, a bunch of antisymmetric tensor
fields also naturally appear in the spectrum and play an important role in
the 
non-perturbative regime, such as D-branes and various dualities, in
superstring theory \cite{Pol}. Thus it is natural to inquire if such 
antisymmetric tensor fields yield a new phenomenon to the still mysterious 
Higgs sector in the Standard Model.
 
In fact, it has been known that when there is a topological term,
antisymmetric 
tensor fields (including gauge field) exhibit an ingenious mass generation 
mechanism, which we call, the 'topological Higgs mechanism' in the sense that 
antisymmetric tensor fields acquire masses and spins without breaking the
local
gauge invariance explicitly. (Of course, in this case, unlike the
conventional 
Higgs mechanism, we do not have the Higgs particle in general, though.)
This interesting mass generation mechanism is first found in the references
\cite{Cremmer, Aurilia, Govindarajan} and then examined in detail within
the framework
of three dimensional gauge theory with Chern-Simons term \cite{Schonfeld,
Deser}.
Afterwards, this three dimensional topological Higgs mechanism has been
generalized 
to an arbitrary higher dimension in cases of both abelian \cite{Minahan,
Oda1, Allen}
and non-abelian gauge theories \cite{Oda2}. More recently, a new type of 
the topological massive nonabelian gauge theories with the usual Yang-Mills
kinetic term has been constructed \cite{Lahiri, Hwang, Landim, Harikumar}.

In this paper, we would like to study the mass generation mechanism in 
the abelian gauge theories \cite{Minahan, Oda1, Allen} and the nonabelian
gauge theories \cite{Lahiri, Hwang, Landim, Harikumar} with the usual Higgs
potential in addition to a topological term \footnote{A preliminary
report has been published \cite{Oda}.}. Since in the Weinberg-Salam
theory the Higgs doublets plus their Yukawa couplings are one of the
key ingredients with great experimental success, we do not want to dismiss 
the framework of the conventional Higgs mechanism. Instead, we wish to
consider how mass of gauge fields is changed and whether the framework of
the conventional Higgs mechanism receives any modification 
if the Higgs potential coexists with a topological 
term. 

Actually, we will discover two novel features in our theory. One feature is 
that gauge fields become massive by 'eating' simultaneously both the 
Nambu-Goldstone boson and a higher-rank tensor field, and instead
a new massless scalar is 'vomitted' in the physical spectrum. 
Consequently, the number of the physical degrees of freedom remains 
unchanged before and after the spontaneous symmetry breakdown. The
other feature is that mass of gauge fields is shifted because of
the topological term and the Higgs potential in a nontrivial
manner. 
Incidentally, about ten years ago, Yahikozawa and the present author have 
studied such a model in the case of the abelian group, but it is a pity 
that there is some misleading statement about the counting of the degrees 
of freedom \cite{Oda3}.
 
This present paper is organized as follows. In section 2, we wish to
correct our previous misleading statement with respect to the counting
of physical degrees of freedom in the abelian model.  In section 3,
we generalize the model to the nonabelian gauge theories. In section 4,
we carry out the manifestly covariant quantization in three and four
space-time dimensions. The final section is devoted to conclusion.

\rm
\section{Abelian gauge theories}

Let us start by reviewing the 'topological Higgs mechanism' in abelian 
gauge theories in an arbitrary space-time dimension \cite{Minahan}:
\begin{eqnarray}
S = - \frac{1}{2} \int dA \wedge \ast dA - \frac{1}{2} \int dB 
\wedge \ast dB + \mu \int A \wedge dB,
\label{1}
\end{eqnarray}
where $A$ and $B$ are respectively an $n$-form and a ($D-n-1$)-form
in $D$-dimensional space-time, "$\ast$" is the Hodge dual operator, 
and $\wedge$ is the Cartan's wedge product, which we will omit henceforth 
for simplicity. In this and next sections, we consider a space-time
with the Euclidean metric signature $(+,+,+, \cdots,+)$, since in this
metric signature many of the factors of $-1$ do not appear in various
equations.  

The equations of motion are easily derived to
\begin{eqnarray}
d \ast d A + (-1)^n \mu d B &=& 0, \nn\\
d \ast d B + (-1)^{n(D-1)+1} \mu d A &=& 0.
\label{2}
\end{eqnarray}
{}From these equations of motion, we obtain
\begin{eqnarray}
( \Delta - \mu^2 ) d A &=& 0, \nn\\
( \Delta - \mu^2 ) d B &=& 0,
\label{3}
\end{eqnarray}
where $\Delta$ is the Laplace-Beltrami operator.
Note that these equations are equations of motion for the transverse
components of $A$ and $B$ fields, and they clearly represent that
two massless antisymmetric tensor fields $A$ and $B$ have gained
the same mass $\mu$ through the topological term $\mu \int A \wedge dB$,
whose phenomenon we call the "topological Higgs mechanism".
It is worthwhile to count the degrees of freedom of physical states
before and after the topological Higgs mechanism occurs. Originally,
we have two massless fields, so the total number of the degrees of freedom
is given by ${D-2 \choose n} + {D-2 \choose D-n-1}$, which is rewritten
as 
\begin{eqnarray}
{D-2 \choose n} + {D-2 \choose D-n-1} &=& {D-1 \choose n} \nn\\
&=& {D-1 \choose D-n-1}.
\label{4}
\end{eqnarray}
This equation clearly indicates that the massless $n$-form field $A$ 
has become massive by 'eating' the massless ($D-n-1$)-form field $B$, 
and vice versa \footnote{Recently, this formalism is utilized for
localizing gauge fields on a brane in brane world \cite{Oda0}.}.

We now turn to the Higgs mechanism in the presence of a topological
term in the abelian gauge theory. To do that, we introduce the conventional 
Higgs potential for only $A$ field, for which we have to restrict $A$ 
field to be a 1-form since the exterior derivative couples with 
only 1-form to make the covariant derivative. Then we have an action 
\cite{Oda3}
\begin{eqnarray}
S &=& - \frac{1}{2} \int dA \ast dA - \frac{1}{2} \int dB \ast dB 
+ \mu \int A dB  \nn\\
&+& \int \Big[ (D \phi)^\dagger \ast D \phi 
- \lambda (|\phi|^2 - \frac{1}{2} v^2) \ast (|\phi|^2 - \frac{1}{2} 
v^2) \Big],
\label{5}
\end{eqnarray}
where $B$ is now a ($D-2$)-form field, and $\phi$ and $v$ are respectively a
0-form complex field and a real number. The covariant derivative $D$ is
defined in a usual way as
\begin{eqnarray}
D \phi &=& d \phi - i g A \phi, \nn\\
(D \phi)^\dagger &=& d \phi^\dagger + i g A \phi^\dagger.
\label{6}
\end{eqnarray}
The gauge transformations are given by
\begin{eqnarray}
\phi(x) &\rightarrow& \phi'(x) = \e^{- i \alpha(x)} \phi(x), \nn\\
A(x) &\rightarrow& A'(x) = A(x) - \frac{1}{g} d \alpha(x), \nn\\
B(x) &\rightarrow& B'(x) = B(x) - d \beta(x),
\label{7}
\end{eqnarray}
where $\alpha(x)$ and $\beta(x)$ are a 0-form and a ($D-3$)-form
gauge parameters, respectively. And $g$ denotes a $U(1)$ gauge
coupling constant. Note that there are still off-shell reducible
symmetries for $B$ field when $D > 3$. 

The minimum of the potential is achieved at 
\begin{eqnarray}
|\phi| = \frac{v}{\sqrt{2}},
\label{8}
\end{eqnarray}
which means that the field operator $\phi$ develops a vacuum
expectation value $|<\phi>| = \frac{v}{\sqrt{2}}$. If we write $\phi$
in terms of two real scalar fields $\phi_1$ and $\phi_2$ as
\begin{eqnarray}
\phi = \frac{1}{\sqrt{2}} ( \phi_1 + i \phi_2 ),
\label{9}
\end{eqnarray}
we can select 
\begin{eqnarray}
<\phi_1> = v,  \ <\phi_2> = 0.
\label{10}
\end{eqnarray}
With the shifted fields
\begin{eqnarray}
\phi'_1 = \phi_1 - v,  \  \phi'_2 = \phi_2,
\label{11}
\end{eqnarray}
we have 
\begin{eqnarray}
<\phi'_1> = <\phi'_2> = 0.
\label{12}
\end{eqnarray}
Note that $\phi'_2$ corresponds to the massless Goldstone
boson.  At this stage, let us take the unitary gauge to remove
the mixing term between $A$ and $\phi'_2$ in the action:
\begin{eqnarray}
\phi^u(x) &=& \e^{-i \frac{1}{v} \xi(x)} \phi(x) = \frac{1}{\sqrt{2}} 
( v + \eta(x) ), \nn\\
G_\mu(x) &=&  A_\mu(x) - \frac{1}{gv} \partial_\mu \xi(x).
\label{13}
\end{eqnarray}
In this gauge condition, $\xi(x)$ and $\eta(x)$ correspond to
$\phi'_2(x)$ and $\phi'_1(x)$, respectively. Also note that the
unitary gauge corresponds to the gauge transformation with 
a fixed gauge parameter $\alpha(x) = \frac{1}{v} \xi(x)$. Then, the 
action (\ref{5}) reduces to the form
\begin{eqnarray}
S &=& - \frac{1}{2} \int dG \ast dG + \frac{1}{2} (g v)^2 \int G \ast G
- \frac{1}{2} \int dB \ast dB \nn\\
&+& \mu \int G dB + \int \Big[ \frac{1}{2} d\eta \ast d\eta - \frac{1}{2} 
(\sqrt{2 \lambda} v)^2 \eta \ast \eta \Big] \nn\\
&+& \int \Big[ \frac{1}{2} g^2 \eta (2 v + \eta) G \ast G 
- \lambda (v \eta \ast \eta^2 + \frac{1}{4} \eta^2 \ast \eta^2) \Big].
\label{14}
\end{eqnarray}
{}From the above action, we can easily read off that the Higgs field 
$\eta(x)$ becomes a massive field with mass $\sqrt{2 \lambda} v$ and 
the would-be Goldstone boson $\xi(x)$ is absorbed into the gauge field 
$G_\mu$, as in the conventional Higgs mechanism without a topological 
term.

In order to clarify the mass generation mechanism for the gauge
field $G$ and the antisymmetric tensor field $B$, it is sufficient to
consider only the quadratic terms with respect to fields in the
action (\ref{14}). In other words, neglecting the interaction terms
in the action (\ref{14}), we  derive the equations of motion for $G$ 
and $B$ fields whose concrete expressions are given by
\begin{eqnarray}
- d \ast dG + (gv)^2 \ast G + \mu dB &=& 0, \\
(-)^D d \ast dB + \mu dG &=& 0.
\label{15,16}
\end{eqnarray}
{}From these equations, we can obtain 
\begin{eqnarray}
\Big[ \Delta - \mu^2 - (gv)^2 \Big] dG &=& 0, \\
\Big[ \Delta - \mu^2 - (gv)^2 \Big] \delta dB &=& 0, \\
\delta G &=& 0,
\label{17,18,19}
\end{eqnarray}
where $\delta$ denotes the adjoint operator. Eqs. (17) and (18) reveal
that fields $G$ and $B$ have become massive fields with the same mass 
$\sqrt{\mu^2 + (gv)^2}$ through the conventional and topological Higgs
mechanisms. Also note that since Eq. (19) holds only when $g v \ne 0$, 
the existence of Eq. (19) reflects a characteristic feature in the case 
at hand.

Now let us consider how the conventional and topological Higgs mechanisms
have worked. For this, it is useful to count the physical degrees of freedom
in the present model. Before spontaneous symmetry breaking, we have two
real scalar fields $\phi_1$ and $\phi_2$, and two massless fields $A_\mu$
and $B_{\mu_1 \cdots \mu_{D-2}}$. The total number of the degrees of freedom
is
\begin{eqnarray}
2 + {D-2 \choose 1} + {D-2 \choose D-2} = D + 1.
\label{20}
\end{eqnarray}
On the other hand, after the symmetry breaking, we have one real scalar
field and one massive field $G_\mu$ (or equivalently, $B_{\mu_1 \cdots 
\mu_{D-2}}$), so it appears that the total number of the degrees of 
freedom is now given by
\begin{eqnarray}
1 + {D-1 \choose 1} = 1 + {D-1 \choose D-2} = D.
\label{21}
\end{eqnarray}
Here, however, we encounter a mismatch of one degree of freedom before and
after
spontaneous symmetry breaking, which is precisely a question raised in 
our previous paper \cite{Oda3}. It is a pity that we have proposed
a misleading answer to this question in the previous paper, so
we wish to give a correct answer in this paper.

To find a missing one physical degree of freedom, we have to
return to the original equations of motion (15) and (16). From
Eq. (16), we have
\begin{eqnarray}
\ast dB + (-)^D \mu G = d\Lambda,
\label{22}
\end{eqnarray}
where $\Lambda$ is a 0-form. Then, using (19) we obtain
\begin{eqnarray}
\Delta \Lambda = 0,
\label{23}
\end{eqnarray}
which means that $\Lambda$ is a massless real scalar field that
we have sought. 
Note that $\Lambda$ is not the Goldstone boson $\xi$ but a new boson 
whose fact can be understood by comparing (13) with (22).
By counting this one degree of freedom, we have
$D + 1$ degrees of freedom after the symmetry breaking which
coincides with the number of the degrees of freedom before the 
symmetry breaking.
Roughly speaking, $G_\mu$ (or $B_{\mu_1 \cdots \mu_{D-2}}$) has become 
massive by 'eating' both the Nambu-Goldstone boson $\xi$ and 
$B_{\mu_1 \cdots \mu_{D-2}}$ (or $G_\mu$), but it has eaten too much
more than its capacity! In consequence, a new massless scalar field has 
been vomitted in the physical spectrum.
We will see in the next section that this interesting new phenomenon 
with respect to the Higgs mechanism in the presence of a topological term 
can be also generalized to the case of the nonabelian gauge theory.

\rm
\section{Nonabelian gauge theories}

We now turn to nonabelian theories. Let us start by reviewing the 
topologically massive nonagelian gauge theories \cite{Lahiri, Hwang, Landim, 
Harikumar}.
The action reads
\begin{eqnarray}
S = \int Tr \Big[ - \frac{1}{2} F \ast F - \frac{1}{2} {\cal{H}} 
\ast {\cal{H}} + \mu B F \Big],
\label{24}
\end{eqnarray}
where we use the following definitions and notations: $F = dA + g A^2, 
H = DB = dB + g [A, B], {\cal{H}} = H + g[F, V]$ and the square 
bracket denotes the graded bracket $[P, Q] = P \wedge Q - (-1)^{|P||Q|} 
Q \wedge P$. And $A$, $B$ and $V$ are respectively a 1-form, a ($D-2$)-form,
a ($D-3$)-form. All the fields are Lie group valued, for instance,
$A = A^a T^a$ where $T^a$ are the generators.
This action is invariant under the gauge transformations
\begin{eqnarray}
\delta A &=& D \theta = d\theta + g [A, \theta], \nn\\
\delta B &=& D \Omega + g [B, \theta], \nn\\
\delta V &=& - \Omega + g [V, \theta], 
\label{25}
\end{eqnarray}
where $\theta$ and $\Omega$ are a 0-form and a ($D-3$)-form gauge
parameters. Recall that a new field strength ${\cal{H}}$ together with
an auxiliary field $V$ has been introduced to compensate for the 
non-invariance of the usual kinetic term $Tr H^2$ under the tensor 
gauge transformations associated with $B$ \cite{Thierry}. From now on, 
we shall set a coupling constant $g$ to be 1 for simplicity 
since we can easily recover it whenever we want.

The equations of motion take the forms
\begin{eqnarray}
D \ast F &=& -[B, \ast {\cal{H}}] + D [V, \ast {\cal{H}}] + \mu DB, \nn\\
D \ast {\cal{H}} &=& (-1)^{D-1} \mu F, \nn\\
{}[F, \ast {\cal{H}}] &=& 0. 
\label{26}
\end{eqnarray}
{}From these equations, we can derive the following equations
\begin{eqnarray}
D \ast D \ast F + (-1)^D \mu^2 F &=& - \mu D \ast [F, V] - D \ast 
[B, \ast {\cal{H}}] + D \ast D [V, \ast {\cal{H}}], \nn\\
D \ast D \ast {\cal{H}} + (-1)^D \mu^2 {\cal{H}} &=& -(-1)^D \mu 
\Big( - \mu [F, V] - [B, \ast {\cal{H}}] + D [V, \ast {\cal{H}}]
\Big). 
\label{27}
\end{eqnarray}
A linear approximation for fields in (27) leads to equations
\begin{eqnarray}
( \Delta - \mu^2 ) dA &=& 0, \nn\\
( \Delta - \mu^2 ) dB &=& 0,
\label{28}
\end{eqnarray}
which imply that the fields $A$ and $B$ become massive by the 
topological Higgs mechanism as in the abelian gauge theories.

Next let us couple the Higgs potential to the model.
For simplicity and definiteness, we shall take the gauge group
to be $G = SU(2)$, with generators $T^i$ satisfying
\begin{eqnarray}
[ T^i, T^j] &=& i \varepsilon^{ijk} T^k, \nn\\
Tr(T^i T^j) &=& \frac{1}{2} \delta^{ij}, \nn\\
T^i &=& \frac{1}{2} \tau^i,
\label{29}
\end{eqnarray}
where $\tau^i$ are the Pauli matrices.
Then, the action is given by
\begin{eqnarray}
S &=& \int Tr \Big[ - F \ast F - {\cal{H}} \ast {\cal{H}} 
+ 2 \mu B F \Big] \nn\\
&+& \int \Big[ (D \phi)^\dagger \ast D \phi 
- \lambda (|\phi|^2 - \frac{1}{2} v^2) \ast (|\phi|^2 - \frac{1}{2} 
v^2) \Big],
\label{30}
\end{eqnarray}
where $F^i_{\mu\nu} = \partial_\mu A^i_\nu - \partial_\nu A^i_\mu
+ \varepsilon^{ijk} A^j_\mu A^k_\nu$ and $D_\mu \phi = ( \partial_\mu
- i \frac{1}{2} \tau^i A^i_\mu) \phi$.

As in the abelian theories, let us take the unitary gauge given by
\begin{eqnarray}
\phi^u(x) &=& U(x) \phi(x) = {0 \choose \frac{1}{\sqrt{2}} 
( v + \eta(x) )}, \nn\\
G(x) &=&  U(x) A(x) U(x)^{-1} + i U(x) dU(x)^{-1}, \nn\\
B'(x) &=&  U(x) B(x) U(x)^{-1}, \nn\\
V'(x) &=&  U(x) V(x) U(x)^{-1},
\label{31}
\end{eqnarray}
where we have defined as $U(x) = \e^{-i \frac{1}{v} \tau^i \xi^i(x)}$.
It then turns out that the action (\ref{30}) reduces to the form
\begin{eqnarray}
S &=& \int Tr \Big[ - F \ast F - {\cal{H}} \ast {\cal{H}} 
+ 2 \mu B F \Big] \nn\\
&+& \int \Big[ (D \phi^u)^\dagger \ast D \phi^u - 
\lambda (v \eta + \frac{1}{2} \eta^2) \ast (v \eta + \frac{1}{2} \eta^2)
\Big],
\label{32}
\end{eqnarray}
where we have rewritten $B'$ and $V'$ as $B$ and $V$, respectively.
Namely, we now have the expressions like $F = dG + G^2$, $D \phi^u 
= d \phi^u - i G \phi^u$. In order to study the mass generation mechanism,
it is sufficient to examine only the quadratic action, which is given by
\begin{eqnarray}
S_0 &=& \int Tr \Big[ - dG \ast dG - dB \ast dB  + 2 \mu B dG \Big] \nn\\
&+& \int \Big[ \frac{1}{2} d\eta \ast d\eta + \frac{1}{2} 
\Big( \frac{gv}{2} \Big)^2
G^i \ast G^i - \lambda v^2 \eta \ast \eta \Big].
\label{33}
\end{eqnarray}
Here we have recovered the coupling constant $g$.
{}From this action (\ref{33}), it is easy to obtain the equations of motion 
\begin{eqnarray}
- d \ast dG^i + \Big( \frac{gv}{2} \Big)^2 \ast G^i + \mu dB^i &=& 0, \\
(-)^D d \ast dB^i + \mu dG^i &=& 0, \\
d \ast d\eta + (\sqrt{2 \lambda} v)^2 \ast \eta &=& 0.
\label{34,35,36}
\end{eqnarray}
{}From these equations, we can obtain 
\begin{eqnarray}
\Big[ \Delta - \mu^2 - \Big(\frac{gv}{2}\Big)^2 \Big] dG^i &=& 0, \\
\Big[ \Delta - \mu^2 - \Big(\frac{gv}{2}\Big)^2 \Big] \delta dB^i &=& 0, \\
\delta G^i &=& 0, \\
\Big[ \Delta - (\sqrt{2 \lambda} v)^2 \Big] \eta &=& 0.
\label{37,38,39,40}
\end{eqnarray}
Eq. (40) shows that the field $\eta$ is indeed the Higgs particle with mass 
$\sqrt{2 \lambda} v$ as in the conventional Higgs mechanism. Also note that 
Eqs. (37)-(39) are the same equations as Eqs. (17)-(19) except the $SU(2)$
index $i$ and the replacement $gv \rightarrow \frac{gv}{2}$, 
so we can show that a completely similar mass generation
mechanism to that in the abelian case occurs also in this
case. Thus the original $SU(2)$ gauge symmetry is completely broken and
all gauge fields (or antisymmetric tensor field) acquire the same mass 
$\sqrt{\mu^2 + (\frac{gv}{2})^2}$ via the conventional and topological Higgs
mechanisms. At the same time, we have a massless scalar with an $SU(2)$
index.

In this case as well, we can check the coincidence of the physical 
degrees of freedom before and after the symmetry breaking as follows.
Before the symmetry breakdown, 
\begin{eqnarray}
4 + 3 \times {D-2 \choose 1} + 3 \times {D-2 \choose D-2} = 3D + 1,
\label{41}
\end{eqnarray}
where 4 is the number of the physical degrees of freedom associated with
the complex Higgs doublet ($\phi$), and 3 appearing in the second and
third terms comes from $SU(2)$. On the other hand, after the symmetry
breakdown
\begin{eqnarray}
1 + 3 + 3 \times {D-1 \choose 1} = 3D + 1,
\label{42}
\end{eqnarray}
where 1 and 3 in the first and second terms equal to the numbers of 
the physical degrees of freedom of the real physical Higgs field
($\eta$) and the new massless field ($\Lambda^i$), respectively.

Before closing this section, we wish to address a few topics related 
to the mechanism clarified in the present paper.
Firstly, we would like to consider how our model sheds some light on 
the Standard Model. For definiteness, let us consider the Weinberg-Salam
model on the basis of $SU(2)_L \times U(1)_Y$ even if it is easy
to extend the model at hand to the Standard Model based on $SU(3)_C \times
SU(2)_L \times U(1)_Y$. 
Note that we can construct a Weinberg-Salam model with topological terms
by unifying the abelian model treated in the previous section and
the $SU(2)$ nonabelian model in this section in an 
$SU(2)_L \times U(1)_Y$-invariant way. Then, we can observe the following
facts: In the conventional Weinberg-Salam model, mass of weak bosons is given 
by $M_W = \frac{gv}{2}$, whereas in our model it is given by 
$m_W = \sqrt{\mu^2 + (\frac{gv}{2})^2}$. Similarly, in our model mass of $Z$ 
boson receives a contribution from a topological term. Concerning the Higgs 
particle, we have the same mass $\sqrt{2 \lambda} v$ in both the models.
Fermion 
masses are also the same in both the models. Thus, we can conclude that
compared 
to the conventional Standard Model, in our model with topological terms we
can 
in general introduce additional parameters stemming from topological terms in 
the mass formulas of gauge bosons without violating the local gauge
symmetries 
explicitly and changing the overall structure of the Standard Model. 
Turning this analysis around, provided that experiment would predict 
$\mu \approx 0$ in future 
it seems that we need to propose some mechanism to suppress the
contribution to 
mass of gauge bosons, since there is $\it{a \ priori}$ no local symmetry
prohibiting
the appearance of topological terms. Moreover, our present model predicts the 
existence of a new massless scalar, which is in a sharp contrast to the 
Standard Model where there is no such a massless boson. These distinct
features in the model at hand will be testable by future experiments.

Secondly, as mentioned in the introduction, many of antisymmetric tensor
fields
naturally appear in the spectrum in superstring theory. For instance,
the action of a supersymmetric $D3$-brane in Type IIB superstring
theory includes topological terms among antisymmetric tensor fields in the 
Wess-Zumino term as well as their kinetic terms in the Born-Infeld action
\cite{Cederwall, Bergshoeff, Aganagic, Kimura1, Kimura2}. 
Hence, it is expected that our model would have some implications in the 
non-perturbative regime of superstring theory. We wish to stress again that 
antisymmetric tensor fields and their topological coupling play a critical
role 
in string dualities \cite{Pol}.

Thirdly, as another implication to superstring theory, note that we have
the term 
$\int_{M_{10}} B \wedge X_8$ in the effective action of superstring theory for
the Green-Schwarz anomaly cancellation. This term yields a topological term
upon compactification to four dimensions. If the Higgs potential appears 
in addition to the topological term via a suitable compactification, the mass 
generation mechanism discussed in the paper would work nicely \cite{Oda3}.

\rm
\section{Manifestly covariant quantization}

In this section, we wish to perform the covariant quantization
of the model considered in the previous section. One of aims of
the quantization is to make certain that the counting of
the physical degrees of freedom done above is really correct.
For instance, if a new massless scalar field $\Lambda$ were
a unphysical or zero norm state, the counting of $\Lambda$ as
the physical degree of freedom would be invalidated. Actually,
we know that there are many unphysical or zero norm states
in topologically massive gauge theories without a mass term
\cite{Naka, Imai, Kimura, Oda1}, so it is worthwhile to carry out the 
quantization of the present model to examine the physical contents in 
detail.

However, it is too complicated to carry out a complete quantization, in
particular, of the nonabelian gauge theories, so we shall limit
ourselves to only the free kinetic terms after the symmetry breakdown,
those are, topological massive gauge theories with a mass term.
Note that it is sufficient to quantize such the theories for seeing
the physical contents. Also note that the nonabelian gauge theory
under a such situation shares the same free fields as in the abelian
gauge theory up to the gauge group index, so we have only to
take account of the abelian theory. Moreover, we shall consider 
the cases of three and four space-time dimensions explicitly, since 
the former case is important for condensed matter physics and is of 
the simplest form whereas the latter case is the most phenomenologically 
interesting one for particle physics.  
The case of a general space-time dimension can be 
treated in a perfectly analogous manner though reducible symmetries get
increased as the number of space-time dimensions becomes larger. 

In this section, we employ not the differential forms but the 
coordinate-dependent component forms, and pick up the Minkowski metric 
$\eta_{\mu\nu} = diag(+1, -1, \cdots, -1)$. 
(Note that we have changed the metric 
signature compared to the previous sections in order to conform with
the notation taken in our paper \cite{Oda1}.) 
The Greek and Latin indices denote the Lorentz and space ones, respectively 
and summation over repeated indices is understood, so the d'Alembertian 
operator, for instance, is defined as $\Box = \partial^\mu \partial_\mu 
= \partial_0^2 - \partial_i^2$.

First of all, let us consider the model in three dimensional flat
Minkowski space-time. The action with which we start is of the form:
\begin{eqnarray}
S &=& \int d^3 x \ \Big[ - \frac{1}{4} (\partial_\mu B_\nu
- \partial_\nu B_\mu)^2 - \frac{1}{4} (\partial_\mu G_\nu
- \partial_\nu G_\mu)^2 + \frac{1}{2} m^2 G_\mu G^\mu  \nn\\
&+& \mu \ \varepsilon^{\mu\nu\rho} B_\mu \partial_\nu G_\rho 
+ B_\mu \partial^\mu N \Big],
\label{43}
\end{eqnarray}
where $\varepsilon_{012} = \varepsilon^{012} = 1$ and we have selected 
the Landau gauge as a gauge condition of the gauge symmetry
$B_\mu \rightarrow B_\mu + \partial_\mu \lambda$. 
This action in essence corresponds to the one (14) obtained after the 
spontaneous symmetry breakdown except that we have neglected the physical
Higgs scalar $\eta(x)$ and its interaction terms. Moreover, we have
defined $m \equiv g v$ for simplicity. The action (43)
leads to the following field equations:
\begin{eqnarray}
&{}& \Box G_\mu - \partial_\mu \partial^\nu G_\nu + m^2 G_\mu
+  \mu \ \varepsilon_{\mu\nu\rho} \partial^\nu B^\rho = 0, \\
&{}& \Box B_\mu - \partial_\mu \partial^\nu B_\nu 
+ \mu \ \varepsilon_{\mu\nu\rho} \partial^\nu G^\rho + 
\partial_\mu N = 0, \\
&{}& \partial^\mu B_\mu = 0.
\label{44,45,46}
\end{eqnarray}
Of course, Eqs. (44) and (45), respectively, correspond to (15) and (16) 
with the auxiliary field $N(x)$ in three dimensions.
Taking the divergence of (44) and (45), we have
\begin{eqnarray}
\partial^\mu G_\mu &=& 0, \\
\Box N &=& 0. 
\label{47,48}
\end{eqnarray}
Applying $\Box$ to (44) and $\Box^2$ to (45), we obtain
\begin{eqnarray}
\Box ( \Box + \mu^2 + m^2 ) G_\mu &=& 0, \\
\Box^2 ( \Box + \mu^2 + m^2 ) B_\mu &=& 0. 
\label{49,50}
\end{eqnarray}

In order to separate massless and massive modes, we define
the fields $S_\mu$ and $V_\mu$ as
\begin{eqnarray}
S_\mu &=& \frac{\mu m}{\mu^2 + m^2} \Big( G_\mu - \frac{1}{\mu}
\varepsilon_{\mu\nu\rho} \partial^\nu B^\rho \Big), \\
V_\mu &=& \frac{1}{(\mu^2 + m^2)^{\frac{3}{2}}} 
\varepsilon_{\mu\nu\rho} \partial^\nu \Box G^\rho,
\label{51,52}
\end{eqnarray}
where we should note that $S_\mu$ precisely corresponds to 
$\partial_\mu \Lambda$ with $\Lambda$ being a new massless scalar
boson vomitted as a result of the spontaneous symmetry breakdown.
Indeed, from the field equations we can show that $S_\mu$ is a
massless field whereas $V_\mu$ is a massive field
\begin{eqnarray}
&{}& \Box S_\mu = \partial^\mu S_\mu = 0, \\
&{}& ( \Box + \mu^2 + m^2 ) V_\mu = \partial^\mu V_\mu = 0. 
\label{53,54}
\end{eqnarray}

We can then derive the following integral identities
\begin{eqnarray}
S_\mu(x) &=& \int d^2 z \ D(x-z) \overleftrightarrow{\partial}_0^z
S_\mu(z), \\
V_\mu(x) &=& \int d^2 z \ \Delta(x-z; \mu^2 + m^2) 
\overleftrightarrow{\partial}_0^z V_\mu(z), \\ 
N (x) &=& \int d^2 z \ D(x-z) \overleftrightarrow{\partial}_0^z N(z), 
\label{55,56,57}
\end{eqnarray}
where $D(x)$ and $\Delta(x; \mu^2)$ denote the invariant $\delta$ 
functions satisfying the relations \cite{Nakanishi1} 
\begin{eqnarray}
\Box D(x) &=& (\Box + \mu^2) \Delta(x; \mu^2) = 0, \nn\\
D(0, \vec{x}) &=& \Delta(0, \vec{x}; \mu^2) = 0, \nn\\
\partial_0 D(0, \vec{x}) &=& \partial_0 \Delta(0, \vec{x}; \mu^2) = 
- \delta(\vec{x}), \nn\\
\Delta(x; \mu^2) &=& \frac{-i}{(2 \pi)^{D-1}} \int d^{D-1}p \
\frac{1}{2 p_0} ( \e^{-ipx} - \e^{ipx} ),
\label{58}
\end{eqnarray}
and we have defined
\begin{eqnarray}
f \overleftrightarrow{\partial} g \equiv (\partial f) g - f \partial g.
\label{59}
\end{eqnarray}
Since it is easy to show that the right-hand sides of Eqs. (55)-(57)
are independent of $z_0$, these integral expressions will be used to
derive the manifestly covariant three dimensional commutation relations
from the equal-time commutation relations.

Let us turn to a canonical quantization of the action (43). The canonical
conjugate momenta corresponding to $B_i$, $G_i$ and $N$ 
are given by
\begin{eqnarray}
p_B^i &=& \frac{\partial \cal{L}}{\partial \dot{B}_i}
= - \dot{B}^i + \partial^i B^0, \nn\\
p_G^i &=& \frac{\partial \cal{L}}{\partial \dot{G}_i}
= - \dot{G}^i + \partial^i G^0 + \mu \varepsilon^{ij} B_j, \nn\\
p_N &=& \frac{\partial \cal{L}}{\partial \dot{N}} = B_0,
\label{60}
\end{eqnarray}
where the dot denotes differentiation with respect to $x^0 \equiv t$.
The following equal-time commutation relations are assumed
\begin{eqnarray}
&{}& [ B_i (\vec{x}, t), p_B^j (\vec{y}, t) ]
= i \delta_i^j \delta (\vec{x} - \vec{y}), \nn\\
&{}& [ G_i (\vec{x}, t), p_G^j (\vec{y}, t) ]
= i \delta_i^j \delta (\vec{x} - \vec{y}), \nn\\
&{}& [ N (\vec{x}, t), p_N (\vec{y}, t) ]
= i \delta (\vec{x} - \vec{y}).
\label{61}
\end{eqnarray}

With the aid of the integral identities (55)-(57), the above
equal-time commutation relations and field equations, we derive
various three dimensional commutation relations whose results are
summarized as
\begin{eqnarray}
&{}& [S_\mu(x), S_\nu(y)] = -i \frac{1}{\mu^2 + m^2}
\partial_\mu \partial_\nu D(x-y), \\
&{}& [S_\mu(x), V_\nu(y)] = 0, \\
&{}& [V_\mu(x), V_\nu(y)] = -i \Big( \eta_{\mu\nu} 
+ \frac{1}{\mu^2 + m^2} \partial_\mu \partial_\nu \Big) \Delta(x-y; \mu^2 +
m^2), \\
&{}& [S_\mu(x), N(y)] = 0, \\
&{}& [V_\mu(x), N(y)] = 0, \\
&{}& [N(x), N(y)] = 0. 
\label{62,63,64,65,66,67}
\end{eqnarray}
As we have obtained the three dimensional commutation relations for
$S_\mu, V_\mu$ and $N$, we can examine the physical contents
of this theory. The physical state condition, the 'Gupta-Bleuler condition'
\cite{Nakanishi2, Lautrup, Nakanishi1}, amounts to 
\begin{eqnarray}
N^{(+)}(x) |phys> = 0,
\label{68}
\end{eqnarray}
where $(+)$ denotes the positive frequency part.

First of all, let us consider the massless field $S_\mu$. In momentum
space, the field equation (53) reduces to the form
\begin{eqnarray}
p^2 s_\mu(p) = p^\mu s_\mu(p) = 0,
\label{69}
\end{eqnarray}
and the three dimensional commutation relations relevant to $S_\mu$,
(62) and (65) are given by
\begin{eqnarray}
&{}& [ s_\mu(p), s_\nu^\dagger(p) ] = \frac{1}{\mu^2 + m^2} 
p_\mu p_\nu, \\
&{}& [ s_\mu(p), n^\dagger(p) ] = 0.
\label{70,71}
\end{eqnarray}
{}From the physical state condition Eq. (68), Eq. (71) means that $s_\mu(p)$ 
is in fact physical. In order to examine the physical contents, it is useful 
to take a Lorentz frame $p_\mu = (p_0 = |p_1|, p_1, 0)$. Then, Eq. (69)
implies
that $s_0(p)$ is not an independent mode but expressed in term of
$s_1(p)$. Moreover, the commutation relations (70) among independent
modes are given by
\begin{eqnarray}
&{}& [ s_1(p), s_1^\dagger(p) ] = \frac{1}{\mu^2 + m^2} p_1^2, \\
&{}& [ s_1(p), s_2^\dagger(p) ] = 0, \\
&{}& [ s_2(p), s_2^\dagger(p) ] = 0. 
\label{72,73,74}
\end{eqnarray}
Eq. (72) shows that $s_1(p)$ is a physical and positive norm mode,
which is nothing but the massless scalar mode found in the previous section!
On the other hand, from (74), $s_2(p)$ is a physical but zero norm mode, so
this mode
is never observed.
It is quite worthwhile to point out that precisely one physical degree of
freedom, that is,  $s_1(p)$, among $s_\mu(p)$ has a positive norm. 

Next, we shall concentrate on a massive field $V_\mu$. 
In momentum space, the field equation (54) becomes 
\begin{eqnarray}
&{}& (p^2 - \mu^2 - m^2) v_\mu(p) = 0, \nn\\
&{}& p^\mu v_\mu(p) = 0,
\label{75}
\end{eqnarray}
and the three dimensional commutation relations, (64) and (66) are given by
\begin{eqnarray}
&{}& [ v_\mu(p), v_\nu^\dagger(p) ] = 
- \Big( \eta_{\mu\nu} - \frac{1}{\mu^2 + m^2} p_\mu p_\nu \Big), \\
&{}& [ v_\mu(p), n^\dagger(p) ] = 0.
\label{76,77}
\end{eqnarray}
Eq. (77) simply means that $v_\mu(p)$ is physical. This time, let us 
take a Lorentz frame $p_\mu = (p_0 = \sqrt{\mu^2 + m^2 + p_1^2}, p_1, 0)$. 
Then, Eq. (75) indicates that $v_0(p) = \frac{p_1}{p_0} v_1(p)$ so $v_0(p)$
is not an independent mode. And the commutation relations (76) become
\begin{eqnarray}
&{}& [ v_1(p), v_1^\dagger(p) ] = \frac{1}{\mu^2 + m^2} p_0^2, \\
&{}& [ v_2(p), v_2^\dagger(p) ] = 1, \\
&{}& [ v_1(p), v_2^\dagger(p) ] = 0. 
\label{78,79,80}
\end{eqnarray}
These commutation relations then mean that only two modes, 
$v_1(p)$ and $v_2(p)$ are not only physical but also positive norm modes
as expected from the arguments in the previous section.
Consequently, we have one massless mode and two massive modes 
as physical degrees of freedom, which exactly coincides with
the result in the previous section. Incidentally, we have carefully
checked by a detailed analysis that the other modes are unphysical or zero 
norm modes.

We are now ready to consider the four dimensional theory.
The line of arguments proceeds as in the three dimensional case.
We begin with the following action in four dimensional flat
Minkowski space-time:
\begin{eqnarray}
S &=& \int d^4 x \ \Big[ \frac{1}{4} \partial_\rho B_{\mu\nu} 
\partial^\rho B^{\mu\nu} - \frac{1}{2} \partial_\rho B_{\mu\nu} 
\partial^\mu B^{\rho\nu} - \frac{1}{4} (\partial_\mu G_\nu
- \partial_\nu G_\mu)^2 + \frac{1}{2} m^2 G_\mu G^\mu  \nn\\
&+& \frac{1}{2} \mu \ \varepsilon^{\mu\nu\rho\sigma} B_{\mu\nu}
\partial_\rho G_\sigma + B_{\mu\nu} \partial^\mu N^\nu 
+ N_\mu \partial^\mu N \Big],
\label{81}
\end{eqnarray}
where the last two terms are the gauge fixing terms. Note that there is
one reducible gauge symmetry for $B_{\mu\nu}$ field, so we need
such two gauge fixing terms. From this action, we obtain
the field equations:
\begin{eqnarray}
&{}& \Box G_\mu - \partial_\mu \partial^\nu G_\nu + m^2 G_\mu
+ \frac{1}{2} \mu \ \varepsilon_{\mu\nu\rho\sigma} \partial^\nu 
B^{\rho\sigma} = 0, \\
&{}& - \frac{1}{2} \Box B_{\mu\nu} - \partial^\rho \partial_{[\mu}
B_{\nu] \rho} + \frac{1}{2} \mu \ \varepsilon_{\mu\nu\rho\sigma} 
\partial^\rho G^\sigma + \partial_{[\mu} N_{\nu]} = 0, \\
&{}& \partial^\mu N_\mu = 0, \\
&{}& \partial_\mu N = \partial^\nu B_{\nu\mu},
\label{82,83,84,85}
\end{eqnarray}
where the square bracket denotes the antisymmetrization with weight 
one, for instance, $\partial_{[\mu} N_{\nu]} = \frac{1}{2}
(\partial_\mu N_\nu - \partial_\nu N_\mu)$.
Taking the divergence of (82), (83) and (85), we obtain
\begin{eqnarray}
\partial^\mu G_\mu &=& 0, \\
\Box N_\mu &=& 0, \\
\Box N &=& 0. 
\label{86,87,88}
\end{eqnarray}
Applying $\Box$ to (82) and $\Box^2$ to (83), we have
\begin{eqnarray}
\Box ( \Box + \mu^2 + m^2 ) G_\mu &=& 0, \\
\Box^2 ( \Box + \mu^2 + m^2 ) B_{\mu\nu} &=& 0. 
\label{89,90}
\end{eqnarray}

In order to separate massless and massive modes, we define
the fields $S_\mu$ and $V_{\mu\nu}$ by
\begin{eqnarray}
S_\mu &=& \frac{\mu m}{\mu^2 + m^2} \Big( G_\mu - \frac{1}{2 \mu}
\varepsilon_{\mu\nu\rho\sigma} \partial^\nu B^{\rho\sigma} \Big), \\
V_{\mu\nu} &=& \frac{1}{(\mu^2 + m^2)^{\frac{3}{2}}} 
\varepsilon_{\mu\nu\rho\sigma} \partial^\rho \Box G^\sigma,
\label{91,92}
\end{eqnarray}
where as in the three dimensional theory $S_\mu$ corresponds to 
$\partial_\mu \Lambda$ with $\Lambda$ being a new massless scalar
boson vomitted as a result of the spontaneous symmetry breakdown.
The field equations tell us that $S_\mu$ is a
massless field whereas $V_{\mu\nu}$ is a massive field
\begin{eqnarray}
&{}& \Box S_\mu = \partial^\mu S_\mu = 0, \\
&{}& ( \Box + \mu^2 + m^2 ) V_{\mu\nu} = \partial^\mu V_{\mu\nu}
= \partial^\nu V_{\mu\nu} = 0. 
\label{93,94}
\end{eqnarray}

We can then derive the following integral identities
\begin{eqnarray}
S_\mu(x) &=& \int d^3 z \ D(x-z) \overleftrightarrow{\partial}_0^z
S_\mu(z), \\
V_{\mu\nu}(x) &=& \int d^3 z \ \Delta(x-z; \mu^2 + m^2) 
\overleftrightarrow{\partial}_0^z V_{\mu\nu}(z), \\ 
N_\mu(x) &=& \int d^3 z \ D(x-z) \overleftrightarrow{\partial}_0^z
N_\mu(z), \\
N (x) &=& \int d^3 z \ D(x-z) \overleftrightarrow{\partial}_0^z N(z).
\label{95,96,97,98}
\end{eqnarray}

Let us move on to a canonical quantization. The canonical
conjugate momenta corresponding to $B_{ij}$, $G_i$, $N_i$, and $N$ 
are calculated as
\begin{eqnarray}
p^{ij} &=& \frac{\partial \cal{L}}{\partial \dot{B}_{ij}}
= \frac{1}{2} \dot{B}^{ij} + \partial^{[i} B^{j] 0}, \nn\\
p_G^i &=& \frac{\partial \cal{L}}{\partial \dot{G}_i}
= - \dot{G}^i + \partial^i G^0 + \frac{1}{2} \mu 
\varepsilon^{ijk} B_{jk}, \nn\\
p_N^i &=& \frac{\partial \cal{L}}{\partial \dot{N}_i} = B^{0i}, \nn\\
p_N &=& \frac{\partial \cal{L}}{\partial \dot{N}} = N_0.
\label{99}
\end{eqnarray}
We set up the following equal-time commutation relations:
\begin{eqnarray}
&{}& [ B_{ij} (\vec{x}, t), p^{kl} (\vec{y}, t) ]
= i \delta_i^{[k} \delta_j^{l]} \delta (\vec{x} - \vec{y}), \nn\\
&{}& [ G_i (\vec{x}, t), p_G^j (\vec{y}, t) ]
= i \delta_i^j \delta (\vec{x} - \vec{y}), \nn\\
&{}& [ N_i (\vec{x}, t), p_N^j (\vec{y}, t) ]
= i \delta_i^j \delta (\vec{x} - \vec{y}), \nn\\
&{}& [ N (\vec{x}, t), p_N (\vec{y}, t) ]
= i \delta (\vec{x} - \vec{y}).
\label{100}
\end{eqnarray}

At this stage, it is straightforward to derive the four dimensional 
commutation relations whose results are of forms
\begin{eqnarray}
&{}& [S_\mu(x), S_\nu(y)] = -i \frac{1}{\mu^2 + m^2}
\partial_\mu \partial_\nu D(x-y), \\
&{}& [S_\mu(x), V_{\nu\rho}(y)] = 0, \\
&{}& [V_{\mu\nu}(x), V_{\rho\sigma}(y)] = i \Big[
\eta_{\mu\rho}\eta_{\nu\sigma}
- \eta_{\mu\sigma}\eta_{\nu\rho} + \frac{1}{\mu^2 + m^2} ( \eta_{\mu\rho}
\partial_\nu \partial_\sigma - \eta_{\mu\sigma} \partial_\nu \partial_\rho \\
&{}& - \eta_{\nu\rho} \partial_\mu \partial_\sigma + \eta_{\nu\sigma}
\partial_\mu 
\partial_\rho ) \Big] \Delta(x-y; \mu^2 + m^2), \\
&{}& [S_\mu(x), N_\nu(y)] = 0, \ [S_\mu(x), N(y)] = 0 \\
&{}& [V_{\mu\nu}(x), N_\rho(y)] = 0, \ [V_{\mu\nu}(x), N(y)] = 0 \\
&{}& [N_\mu(x), N_\nu(y)] = 0, \ [N(x), N(y)] = 0 \\
&{}& [N_\mu(x), N(y)] = i \partial_\mu D(x-y). 
\label{101,102,103,104,105,106,107,108}
\end{eqnarray}
As we have obtained the four dimensional commutation relations for
$S_\mu, V_{\mu\nu}, N_\mu$ and $N$, we can examine the physical contents
of this theory. The physical state condition, the 'Gupta-Bleuler condition',
becomes
\begin{eqnarray}
N_\mu^{(+)}(x) |phys> = N^{(+)}(x) |phys> = 0.
\label{109}
\end{eqnarray}

We shall first consider the massless field $S_\mu$. In momentum
space, as in the case of the three dimensional theory the field equation 
(93) takes the form
\begin{eqnarray}
p^2 s_\mu(p) = p^\mu s_\mu(p) = 0,
\label{110}
\end{eqnarray}
and the four dimensional commutation relations relevant to $S_\mu$
are given by
\begin{eqnarray}
&{}& [ s_\mu(p), s_\nu^\dagger(p) ] = \frac{1}{\mu^2 + m^2} 
p_\mu p_\nu, \\
&{}& [ s_\mu(p), n_\nu^\dagger(p) ] = [ s_\mu(p), n^\dagger(p) ] = 0.
\label{111,112}
\end{eqnarray}
Then Eq. (112) means that $s_\mu(p)$ is physical. Again let us take a 
Lorentz frame $p_\mu = (p_0 = |p_1|, p_1, 0, 0)$. Then, Eq. (110) implies
that $s_0(p)$ is not an independent mode but expressed in term of
$s_1(p)$. Moreover, the commutation relations (111) among independent
modes are given by
\begin{eqnarray}
&{}& [ s_1(p), s_1^\dagger(p) ] = \frac{1}{\mu^2 + m^2} p_1^2, \\
&{}& [ s_2(p), s_2^\dagger(p) ] = [ s_3(p), s_3^\dagger(p) ] = 0, \\
&{}& [ s_1(p), s_2^\dagger(p) ] = [ s_1(p), s_3^\dagger(p) ] = 
[ s_2(p), s_3^\dagger(p) ] = 0, 
\label{113,114,115}
\end{eqnarray}
Eq. (113) shows that $s_1(p)$ is a physical and positive norm mode,
which precisely corresponds to the massless scalar mode found in the 
previous section, whereas $s_2(p)$ and $s_3(p)$ are physical but zero 
norm modes, so these modes are never observed.

Next, we shall turn our attention to a massive field $V_{\mu\nu}$. 
In momentum space, the field equation (94) becomes 
\begin{eqnarray}
&{}& (p^2 - \mu^2 - m^2) v_{\mu\nu}(p) = 0, \nn\\
&{}& p^\mu v_{\mu\nu}(p) = p^\nu v_{\mu\nu}(p) = 0,
\label{116}
\end{eqnarray}
and the four dimensional commutation relations, (103) and (106) are given by
\begin{eqnarray}
&{}& [ v_{\mu\nu}(p), v_{\rho\sigma}^\dagger(p) ] = 
\eta_{\mu\rho}\eta_{\nu\sigma} - \eta_{\mu\sigma}\eta_{\nu\rho} 
- \frac{1}{\mu^2 + m^2} ( \eta_{\mu\rho} p_\nu p_\sigma  \nn\\
&{}& -\eta_{\mu\sigma} p_\nu p_\rho - \eta_{\nu\rho} p_\mu p_\sigma 
+ \eta_{\nu\sigma} p_\mu p_\rho ), \\
&{}& [ v_{\mu\nu}(p), n_\rho^\dagger(p) ] = [ v_{\mu\nu}(p), n^\dagger(p) ] 
= 0.
\label{117,118}
\end{eqnarray}
Eq. (118) expresses that $v_{\mu\nu}(p)$ is physical. For a further
analysis, let us take a Lorentz frame 
$p_\mu = (p_0 = \sqrt{\mu^2 + m^2 + p_1^2}, p_1, 0, 0)$. 
Then, Eq. (116) gives us the relations
\begin{eqnarray}
&{}& v_{01}(p) = 0, \nn\\
&{}& v_{02}(p) = \frac{p_1}{p_0} v_{12}(p), \nn\\
&{}& v_{03}(p) = \frac{p_1}{p_0} v_{13}(p),
\label{119}
\end{eqnarray}
thereby impling that $v_{01}(p)$, $v_{02}(p)$ and $v_{03}(p)$
are not independent modes.
Furthermore, the commutation relations (117) among independent
modes are given by
\begin{eqnarray}
&{}& [ v_{12}(p), v_{12}^\dagger(p) ] = \frac{1}{\mu^2 + m^2} p_0^2, \\
&{}& [ v_{13}(p), v_{13}^\dagger(p) ] = \frac{1}{\mu^2 + m^2} p_0^2, \\
&{}& [ v_{23}(p), v_{23}^\dagger(p) ] = 1, 
\label{120,121,122}
\end{eqnarray}
and the remaining commutation relations are found to be trivially
vanishing. These commutation relations then mean that only three
physical degrees of freedom associated with $v_{12}(p)$, $v_{13}(p)$ and 
$v_{23}(p)$ are not only physical but also positive norm modes
while the other $v_{\mu\nu}(p)$ are physical but zero norm modes
as expected from the arguments in the previous section.
As a result, we have one massless mode and three massive modes 
as physical degrees of freedom, which again coincides with
the result in the previous section. It is in principle straightforward
to extend our analysis in three and four dimensions to higher
dimensions, but as we study higher dimensions, we will encounter
more technical complications owing to increasing reducible
gauge symmetries associated with the antisymmetric tensor gauge
fields.

\rm
\section{Conclusion}

In this paper, we have investigated a physical situation where
the Higgs potential coexists with a topological term. We have
found that the gauge field becomes massive by 'eating' the Nambu-Goldstone
boson and the antisymmetric tensor field, and 'vomits'
a new massless scalar field. Moreover, we have pointed out that 
when our model is extended to a more realistic model such as the 
Weinberg-Salam model and the Standard Model, it gives us some distinct
results, those are, the shift of mass of gauge fields and the presence
of a new massless boson, which would be checked in future by experiment.
Of course, such a mass shift would be very small in such a way as to 
ensure that the mass of weak bosons matches the experimental data.  

Although the experiment might preclude our model and support the
Weinberg-Salam theory in future experiment, it would be necessary 
to propose some mechanism for suppressing the effects coming from a
topological term since there is no symmetry prohibiting the existence
of such a topological term. In this context, we should recall that
the topological term under consideration is manifestly CPT-invariant,
which should be contrasted with 3D Chern-Simons term.
This problem becomes more acute in superstring-inspired phenomenology 
since superstring theory gives rise to many of antisymmetric tensor fields 
with a topological term at low energy.

To make the present model viable, we need to give the proof
of unitarity and renormalizability. The proof of unitarity can
be done by extending the formalism of a canonical quantization in section
4 \cite{Oda4}, but the proof of renormalizability needs much work.

\vs 1
\begin{flushleft}
{\bf Acknowledgement}
\end{flushleft}
We are grateful to A. Lahiri for useful informations
on related references .

\vs 1


\end{document}